\documentstyle[aps,preprint,epsfig]{revtex}

\begin{document}

\tightenlines

\preprint{\parbox[b]{1in}{ \hbox{\tt
PNUTP-03/A01} \hbox{\tt OITS-727} }}

\draft

\title{High Temperature Superfluid and Feshbach Resonance}

\author{Deog Ki Hong$^{1,a}$ and Stephen D.H.~Hsu$^{2,b}$}

\vspace{0.05in}

\address{
$^{a}$Department of Physics, Pusan National University, Pusan
609-735, Korea
\protect\\
$^b$Department of Physics, University of Oregon, Eugene OR
97403-5203
\protect\\   
\vspace{0.05in} {\footnotesize\tt $^{1}$dkhong@pnu.edu,
$^2$hsu@duende.uoregon.edu}}

\vspace{0.1in}

\date{\today}

\maketitle

\begin{abstract}
We study an effective field theory describing cold fermionic atoms
near a Feshbach resonance. The theory gives a unique description
of the dynamics in the limit that the energy of the Feshbach
resonance is tuned to be twice that of the Fermi surface. We show
that in this limit the zero temperature superfluid condensate is
of order the Fermi energy, and obtain a critical temperature $T_C
\simeq 0.43~T_F$
\end{abstract}


\newpage
After the successful achievement of Bose-Einstein condensation in
ultracold atomic gases, a natural question is whether one can
achieve superfluidity in a degenerate fermion gas. Samples of
fermionic alkali atoms have already been cooled below their Fermi
temperature of several hundreds nano-Kelvin \cite{jin-fermi},
where quantum degeneracy phenomena such as Cooper-pairing of atoms
are potentially important~\cite{thomas}.

It was suggested that the superfluidity gap may be made as large
as of order the Fermi energy in the alkali atomic
gases~\cite{holland1,timmermans,griffin}, using a Feshbach
resonance~\cite{feshbach}, at which the scattering length becomes
very large. For weak coupling, the scattering length away from
resonance is usually much smaller than the typical wavelength of
fermions near the Fermi surface, $\left|a\right|\ll k_F^{-1}$ and
the Cooper pairing gap is then given in terms of scattering length
as
\begin{equation}
\Delta\simeq 2E_F e^{-\pi/2k_F\left|a\right|},
\label{scattering}
\end{equation}
where $E_F$ is the Fermi energy and $\hbar k_F=p_F$ is the Fermi
momentum.  The effective strength of attraction is related to the
scattering length as $V=4\pi\hbar^2 an/m$, where $n$ is the number
density and $m$ is the fermion mass.
Therefore, a weakly bound Cooper-pair gap is typically much
smaller than the Fermi energy. However, if the scattering of
fermions occurs at the Feshbach resonance, whose energy is nearly
equal to twice that of the Fermi surface, the effective scattering
length becomes quite large and so does the gap. In this paper we
analyze this enhancement at the Feshbach resonance in detail.

At ultralow temperature, the atomic gas can be described by a
Fermi liquid {\it {\`a} la } Landau, whose (effective) Lagrangian
density is given as, suppressing the spin indices,
\begin{equation}
{\cal L}_{\rm eff}
=\psi^{\dagger}\left(i\partial_t+ {{\nabla}^2\over 2m}\right)
\psi+E_F\psi^{\dagger}\psi+{\cal L}_{\rm int},
\end{equation}
where ${\cal L}_{\rm int}$ describes the interaction of atoms. The
interaction Lagrangian can be expanded in powers of $\psi$ and
$\vec\nabla\psi$,
\begin{equation}
{\cal L}_{\rm int}={g_1\over2}\left(\psi^{\dagger}\psi\right)^2
+g_2\left(\psi^{\dagger}\vec\nabla\psi\right)^2+\cdots
\end{equation}
Near the Fermi surface, interactions with non-Cooper pairing
kinematics, as well as derivative corrections to the Cooper
pairing interaction, are irrelevant and flow to
zero~\cite{Shankar:1994pf}. For scattering of atoms of opposite
momenta, only the leading term is marginal, and, if attractive
($g_1>0$), a gap opens at the Fermi surface, given as
\begin{equation}
\Delta\simeq\, 2E_F\, e^{-2/(g_1\nu_F)},
\end{equation}
where $\nu_F$ is the energy density of states at the Fermi surface.

However, when the resonance scattering occurs at energy near twice
the Fermi energy, $E_R\simeq0$, we have to include the resonance
degrees of freedom in the effective Lagrangian, since such a
resonance state is easily excited and has a long lifetime:
\begin{equation}
\label{resonanceL}
{\cal L}_{\rm eff}\ni \phi^{\dagger}\left(i\partial_t+{\nabla^2\over 2M}
-E_R-i\Gamma\right)\phi+g\left(\phi\,\psi^T i\sigma_2\psi +{\rm
h.c.} \right)+\cdots,
\end{equation}
where $\phi$ denotes the (spin-0) resonance field of mass $M$ and
a decay width $\Gamma$. The ellipsis denotes the higher-order
terms involving $\phi$, $\vec \nabla\phi$, and/or $\psi$,
$\vec\nabla \psi$, which are negligible near the Fermi surface.
Note that the resonance energy $E_R$ is measured from the Fermi
surface. Namely, it is the binding energy minus twice the Fermi
energy.
As the coupling g has dimensions of $[\rm energy]^{-1/2}$
in natural units, we introduce a dimensionless coupling
$\bar{g}$,
\begin{equation}
\bar{g}= \sqrt{2M} g~~~,
\end{equation}
which is simply the Yukawa coupling between the scalar resonance
and the fermionic atom that would appear in a relativistic
formulation of the model. (We can obtain the kinetic energy in
(\ref{resonanceL}) from the usual kinetic term $\partial_\mu \phi
\partial^\mu \phi$ by rescaling the field $\phi$ by
$e^{iMt}/\sqrt{2M}$.) By comparing on-resonance scattering in the
relativistic and non-relativistic formulations, one can see that
$\bar{g}^2 \sim \Gamma / p_F$.

\begin{figure}
\vskip 0.1in
\epsfxsize=2in
\centerline{\epsffile{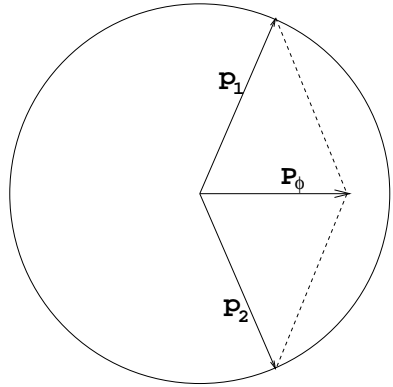}}
\vskip 0.1in
\caption{The resonance decays into two atoms in medium. The circle denotes
the Fermi surface.}
 \label{fig1}
\end{figure}

\begin{figure}
\epsfxsize=3in
\centerline{\epsffile{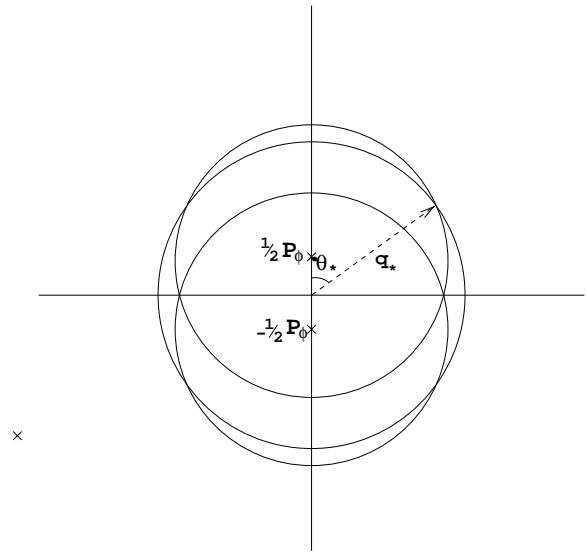}}
\caption{The allowed angles in decay of the resonance.}
 \label{fig2}
\end{figure}

We now consider the decay process of the resonance in medium (Fig.~{\ref{fig1}}):
\begin{equation}
\phi\longrightarrow \psi +\psi.
\end{equation}
The differential decay rate in the rest frame of medium is
\begin{eqnarray}
d\Gamma^*=(2\pi)^4\delta^4(p_1+p_2-p_{\phi}){1\over 2E_{\phi}}
\cdot {m^2\over E_1E_2}\left[1-f(E_1)\right]{d^3p_1\over (2\pi)^3}
\left[1-f(E_2)\right]{d^3p_2\over (2\pi)^3}
\sum_s\left|{\cal M}\right|^2,
\end{eqnarray}
where $E_i=\sqrt{m^2+p_i^2}$, $E_{\phi}=\sqrt{M^2+p_{\phi}^2}$,
and $f(E)=\left[1+e^{-(\bar\mu-E)/T}\right]^{-1}$ with
$\bar\mu=\sqrt{m^2+p_F^2}$. The $T$ matrix in the non-relativistic
limit is $\left|{\cal M}\right|^2\simeq g^2$. Because of the Pauli
blocking, the decay process is restricted by kinematics. For a
two-body decay, the Fermi-Dirac distribution restricts the angle,
$\theta$, between $\vec p_{\phi}$ and $\vec q\equiv(\vec p_1-\vec
p_2)/2$ (see Fig.~\ref{fig2}).

When $p_{\phi}\le 2\sqrt{p_F^2-q_*^2}$ with
$q_*^2=m\hat E_{\phi}-p_{\phi}^2/4$, where
$\hat E_{\phi}=E_R+2E_F+{p_{\phi}^2\over 2M}$,
the decay width near $T\simeq0$ is
\begin{eqnarray}
\Gamma^*={1\over\pi}g^2{m^4\over \hat E_{\phi}^2p_{\phi}}\,
\ln\left[{1+{p_{\phi} q_*\over m\hat E_{\phi}}\cos\theta_*\over
1-{p_{\phi} q_*\over m\hat E_{\phi}}\cos\theta_*}
\right],
\end{eqnarray}
where $\left|\vec q -\vec p_{\phi}/2\right|=p_F$
at $\theta=\theta_*$ and $|\vec q|=q_*$.
When $p_F+|\vec p_{\phi}|/2<q_*$, the angle $\theta$ is not restricted
and $\Gamma^*=\Gamma^*(\theta_*=0)$. Finally, when $p_F^2-p_{\phi}^2/2
>q_*^2$ or $p_{\phi}^2<p_F^2/m$, the resonance does not decay, $\Gamma^*=0$,
as the final states allowed by kinematics are Pauli-blocked.

The decay rate of the resonance in free space can be similarly calculated.
We find that the resonance in free space decays at rest with a rate,
\begin{equation}
\Gamma={{\bar g}^2\over\pi}\sqrt{m{\hat E}_R},
\end{equation}
where $\hat E_R=E_R+2E_F$ is the resonance energy in free space.
According to a recent measurement~\cite{jin}
of a Feshbach resonance between optically
trapped $^{40}{\rm K}$ atoms in the two lowest energy spin-states,
$\left|9/2,-9/2\right>$ and $\left|9/2,-7/2\right>$, the resonance peak occurs
at a magnetic field of $201.5~\pm1.4~{\rm G}$ with a width of
$8.0~\pm1.1~{\rm G}$,
which corresponds to ${\hat E}_R=10^{-6}~{\rm eV}$ and
$\Gamma=4\times10^{-8}~{\rm eV}$.
For experimentally realizable ultracold atoms, $E_F/k_B\sim 600~{\rm nK}$ or
$E_F\sim 6\times 10^{-11}~{\rm eV}$. The dimensionless coupling $\bar{g}$ is
much less than one.

Now let us consider the medium effect on the propagation of the
resonance. Even though the gas of atoms is dilute, the medium
effect is important when the de Broglie wavelength is larger than
the interatomic distance. The self energy of the resonances is
\begin{eqnarray}
i\Sigma(p)=2g^2
\int_q{1\over
(1+i\epsilon)\left({p_0\over2}+q_0\right)
-{\left(\vec p/2+\vec q\right)^2\over 2m}+E_F}\cdot
{1\over
(1+i\epsilon)\left({p_0\over2}-q_0\right)
-{\left(\vec p/2-\vec q\right)^2\over 2m}+E_F}.
\end{eqnarray}
We perform the $q_0$ integration first and rewrite the self energy
as sum of the vacuum part and the matter part from the states in
the Fermi sea:
\begin{eqnarray}
i\Sigma(p)=i\Sigma_{\rm vac}(p)+i\Sigma_{\rm matt}(p),
\end{eqnarray}
where
\begin{eqnarray}
i\Sigma_{\rm vac}(p)=2ig^2\int{d^3q\over (2\pi)^3}
{1\over p_0-\left(\vec p^2/4+\vec q^2-2 p_F^2\right)/2m},
\end{eqnarray}
which is irrelevant for our purpose,
and the matter part becomes
\begin{eqnarray}
i\Sigma_{\rm matt}(p)=-4i\,mg^2
\int{d^3q\over (2\pi)^3}
{\theta\left(p_F-\left|\vec q+{\vec p\over2}\right|\right)+
\theta\left(p_F-\left|\vec q-{\vec p\over2}\right|\right)
\over2p_F^2+2mp_0-\vec p^2/4m-\vec q^2}
\end{eqnarray}
When $p_0\ll E_F,~|\vec p|\ll p_F$,
\begin{equation}
\Sigma_{\rm matt}(p)={g^2\over 2\pi^2}m\,p_F\left[
1+O\left({p_0\over E_F},{|\vec p|\over p_F}
\right)\right].
\end{equation}
We note that the medium gives a finite contribution to the resonance
energy:
\begin{equation}
E_R\longrightarrow E_R^*\equiv E_R-{{\bar{g}}^2 \over 4 \pi^2 v_F} E_F~~~.
\end{equation}

When the resonance energy is large or of order of $E_F$, the
resonance degrees of freedom are irrelevant and can be integrated
out in favor of a four-Fermi interaction. However, if the
resonance energy is tuned to be much smaller than the Fermi
energy, it must be included in the low energy dynamics of the cold
atomic gas. A drastic effect occurs when the resonance energy is
negative (or less than twice the Fermi energy of atoms). Namely,
the resonance cannot decay into two atoms due to the Pauli
blocking. The lifetime is proportional to $1/T$ and becomes
infinity at zero temperature. The system can reduce its energy by
removing fermions from the Fermi sea and storing them as condensed
bosons. In this regime we expect Bose-Einstein condensation of the
resonance \cite{holland1,griffin}.

\begin{figure}
\epsfxsize=3in
\centerline{\epsffile{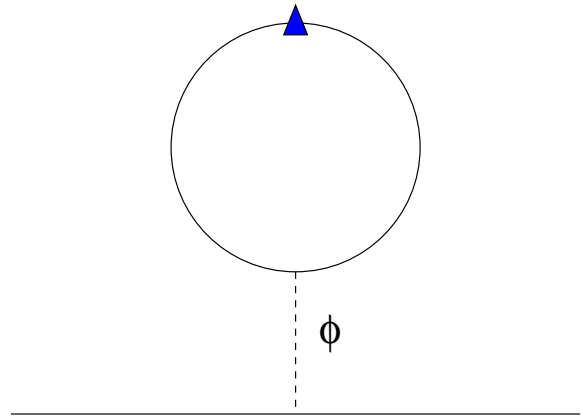}}
\vskip 0.1in
\caption{The dashed line denotes the resonance and the triangle denotes
the superfluid gap.}
 \label{fig3}
\end{figure}

Here, we are most interested in the case of zero detuning, when
the resonance is very close to the Fermi surface, and our
effective theory applies with great accuracy. The Cooper theorem
then tells us that the Fermi surface becomes unstable to
atom-pairing, which is described here as a condensate of
resonances (see Fig.~\ref{fig3}).

We determine the value of the condensate by calculating the
Coleman-Weinberg potential for the resonance~\cite{Coleman:jx},
using a formalism developed for a Fermi surface effective
theory~\cite{Hong:2000tn}.
\begin{eqnarray}
V(\phi)=(E_R+i\Gamma)\phi^{\dagger}\phi+B\phi^{\dagger}\phi
-{2p_F^2\over\pi}\int{{\rm d}E{\rm dl}\over (2\pi)^2}
\ln\left(1+{g^2\phi^{\dagger}\phi\over E^2+v_F^2l^2}\right),
\end{eqnarray}
where $B$ is the counter term and higher orders are suppressed,
as we discuss in more detail below.
Choosing the renormalization condition at $g\phi=\Lambda$ to be
\begin{equation}
\label{rcond} \left.{1\over 2}{{\rm d}^2V\over {\rm d}\phi^2}
\right|_{g\phi=\Lambda}=E_R^*+i\Gamma^*,
\end{equation}
where $E^*_R=E_R-g^2mp_F/(2\pi^2)=E_R-{{\bar{g}}^2 E_F/
(4 \pi^2v_F)}$ and $\Gamma^*$ are the resonance energy and the decay rate
in medium, respectively, we find the real part of the potential to
be
\begin{equation}
V(\phi)=E_R^*\phi^{\dagger}\phi +{p_F^2\over2\pi^2
v_F}g^2\,\phi^{\dagger}\phi\,\ln\left( {g^2\phi^{\dagger}\phi\over
\Lambda^2}\right),
\end{equation}
which has a minimum at
\begin{equation}
\left<\phi\right>={\Lambda\over g}\exp\left(-{1\over2}-{2\pi^2
E_R^*v_F\over {\bar g}^2E_F}\right).
\end{equation}
Taking the renormalization point $\Lambda=E_F$, we find the
superfluidity gap
\begin{equation}
\Delta=E_F \,e^{-1/2- 2\pi^2 E_R^* v_F /{\bar{g}}^2 E_F},
\end{equation}
where $E_R^*$ is the resonance energy in medium measured at the
Fermi momentum. The renormalization condition (\ref{rcond}) resums
the medium corrections to the resonance propagator (i.e. vacuum
polarization due to atom loops). These effects modify the
resonance-induced interaction between atoms
(see Fig.~\ref{fig3}), as well as the definition of zero detuning. If we
tune the resonance energy in medium to be zero,
the gap becomes as large as $\Delta=e^{-1/2}E_F$. This result is a
robust consequence of the effective theory, depending only on the
existence of a resonance near the Fermi surface.

Corrections to our results come from two sources: (1) higher order
graphs within the effective theory, and (2) corrections to the
effective theory itself.

It is straightforward to estimate corrections of type (1), which
are obtained by inserting, e.g., in Fig.~3, an additional
resonance exchange on the internal atom line. The resulting
expression is suppressed by an additional factor of $\bar{g}^2$,
but does not produce additional powers of $k_F^2$ (the area of the
Fermi surface) for kinematical reasons. For this reason, the
Cooper pairing channel already considered gives the dominant
contribution and additional resonance exchange interactions are
negligible.

On the other hand, when the gap is as large as $\Delta =
e^{-1/2}E_F$, some of the pairing atoms are far from the Fermi
surface. The effective theory description there may suffer
corrections of type (2) which are order $\Delta$ over $E_F$, and
might decrease the size of the gap. However, by self-consistency,
the resulting $\Delta$ cannot be much smaller than $E_F$ (perhaps
by an order of magnitude at most)\footnote{This is in agreement
with an earlier result \cite{neutron} using effective field theory
to write the gap directly in terms of the two-particle phase
shift: $\Delta \sim E_F \exp (- \frac{\pi}{2} \cot \delta )$. At a
resonance, $\cot \delta$ goes through zero, so $\Delta$ is of
order $E_F$. A simple estimate using the gap equation obtained
from Fig.~3 also yields $\Delta \sim E_F \exp( {\cal O}(1) )$ when
the scalar exchange is exactly on-resonance.}.

Let us now consider finite temperature effects. Strictly speaking,
the calculation of the critical temperature involves fluctuations
which are far from the Fermi surface, and hence is not fully
described by the effective theory. Also, we must assume that the
resonances maintain thermal equilibrium with the atoms without
being destroyed by the fluctuations. The Coleman-Weinberg
potential becomes
\begin{equation}
V(\phi,T)=E_R\phi^{\dagger}\phi+B\phi^{\dagger}\phi-{2p_F^2\over\pi}k_BT
\sum_{n=-\infty}^{+\infty}\int{{\rm d}l\over2\pi}
\ln\left[1+{g^2\phi^{\dagger}\phi\over
\omega_n^2+v_F^2l^2}\right],
\end{equation}
where the Matsubara frequency $\omega_n=\pi k_BT(2n+1)$.
The minimum occurs at
\begin{equation}
E_R+B=g^2{p_F^2\over \pi^2 v_F}k_BT\sum_{n=-\infty}^{\infty}
\int{{\rm d}l}{1\over \omega_n^2+l^2+g^2\phi^{\dagger}\phi}.
\end{equation}
As the temperature approaches the critical temperature,
the gap vanishes and the minimum therefore should
occur at $\phi=0$, which gives, taking the renormalization point
to be $E_F$,
\begin{equation}
E_R={p_F^2\over \pi^2 v_F}g^2
\int_0^{E_F}{{\rm d}l\over l}\,\tanh\left({l\over 2k_BT_C}\right).
\end{equation}
The critical temperature at which the gap vanishes is therefore
\begin{eqnarray}
T_C={2e^\gamma\over \pi k_{\rm B}}\Delta,
\end{eqnarray}
which is twice the BCS value ($\gamma\simeq0.577$ is the Euler
constant). At the Feshbach resonance in medium, $E_R^*=0$ or
$E_R=g^2mp_F/(2\pi^2)$, we find $T_C=0.68\,E_F/k_B$. This is
slightly larger than the value obtained by Holland {\it et
al.}~\cite{holland1} At the Feshbach resonance the ultracold Fermi
gas becomes a superfluid if it cools below $0.68~T_F$, where $T_F$
is the Fermi temperature.

Because the critical temperature obtained is large (of order
$T_F$), this calculation must be modified to correctly account for
atom number conservation. In other words, we must impose the
condition that the number density of atoms
\begin{equation}
\label{number}
 N_0 = N_F + N_B~~~,
\end{equation}
where $N_0$ is the number density in the absence of the resonance
and $N_F$ and $N_B$ are functions of a chemical potential $\mu$
which may differ from $E_F$. Using the free particle Fermi-Dirac
and Bose-Einstein distributions in $N_{F,B}$ along with the
relation $T_C = .68 \mu$, we obtain $\mu \simeq .63 E_F$, which
suggests that number conservation reduces the critical temperature
by roughly one third. A more sophisticated expression for the
number densities $N_{F,B}$ can be obtained by including the effect
of interactions on the thermodynamic potential \cite{griffin}. We
find that the fluctuation contribution to the number density is
\begin{eqnarray}
N_C\simeq{p_F^2\over 2\pi^3v_F}{\Delta(T)^2\over\mu}\tanh\left({\mu\over T}\right)
-{p_F^2\over2\pi^2 v_F}T\left[
{E_R-E_R^*\over T}-\ln\left({1+e^{(\mu/2-E_R^*)/T}
\over 1+e^{(\mu/2-E_R)/T}}\right)
\right].
\end{eqnarray}
Unless the resonance-atom coupling is taken to be very large (so
large as to render a controlled calculation doubtful), the effect
of interactions on (\ref{number}) is quite small near the critical
temperature.

In summary, we used an effective field theory describing modes
near the Fermi surface to compute the Coleman-Weinberg
(mean-field) potential for the resonance in medium, obtaining a
robust superfluidity gap of $\Delta=e^{-1/2}E_F$ at zero detuning.
We find that the critical temperature for the superfluid phase
transition is $T_C\simeq0.43~T_F$, taking into account the
constraint from conservation of atom number.

\eject

\acknowledgments

We thank Anthony Leggett for discussions. The work of D.K.H.
was supported in part by the KOSEF grant number 1999-2-111-005-5
and also  by the academic research fund of Ministry of Education,
Republic of Korea, Project No. BSRI-99-015-DI0114. The work of
S.H. was supported in part under DOE contract DE-FG06-85ER40224
and by the NSF through the USA-Korea Cooperative Science Program,
9982164.

\end{document}